\documentclass[a4paper,11pt]{article}
\usepackage{pos}
\usepackage{cleveref}

\newcommand{\msbar}{\rm \overline{MS}}
\newcommand{\rmd}{\mathrm{d}}
\newcommand{\rmO}{\mathrm{O}}

\newcommand{\vecp}{\bf  p}
\newcommand{\vecx}{\bf  x}

\newcommand{\simas}[1]{\raisebox{-.1ex}{
            $\stackrel{\small{#1}}{\sim}$}}

\newcommand{\gammahat}{\hat\gamma}

\newcommand{\kl}{k_\mathrm{L}}
\newcommand{\mom}{\mathcal{M}}

\title{Log-enhanced discretization errors in integrated correlation functions}

\author[a,b]{Leonardo Chimirri}
\author[c]{Nikolai Husung}
\author*[a,b]{Rainer Sommer}

\affiliation[a]{Deutsches Elektronen-Synchrotron DESY, \\ 
Platanenallee~6, 15738~Zeuthen, Germany}

\affiliation[b]{
Institut~f\"ur~Physik, Humboldt-Universit\"at~zu~Berlin, \\
 Newtonstr.~15, 12489~Berlin, Germany}

\affiliation[c]{Physics and Astronomy, University of Southampton, \\
Southampton SO17 1BJ, United Kingdom}

\emailAdd{rainer.sommer@desy.de}

\abstract{Integrated time-slice correlation functions $G(t)$ with weights $K(t)$ appear, e.g., in
the moments method to determine $\alpha_s$ from heavy quark correlators,
in the muon g-2 determination or in the determination of smoothed spectral
functions. 

For the (leading-order-)normalised moment $R_4$ of the pseudo-scalar correlator 
we have non-perturbative results down to $a=10^{-2}$ fm and for masses, $m$, of the order of the charm 
mass in the quenched approximation. A significant bending of $R_4$ as a function of $a^2$ is observed at small lattice
spacings.

Starting from the Symanzik expansion 
of the integrand we derive the asymptotic convergence of the integral at small lattice spacing in the free theory and prove
that the short distance part of the integral leads to $\log(a)$-enhanced 
discretisation errors when $G(t)K(t) \simas{t\to 0}\, t $ for small $t$. 
In the interacting theory an unknown, 
function $K(a\Lambda)$ appears. 

For the $R_4$-case, we  modify the observable to improve the short distance behavior and demonstrate that it results in a very smooth continuum limit. The strong coupling and the $\Lambda$-parameter can then be extracted. In general, and in particular for $g-2$, the short distance part of the integral should be determined by perturbation theory. The (dominating) rest can then be obtained by the controlled continuum limit of the lattice computation.
}

\FullConference{%
  The 39th International Symposium on Lattice Field Theory (Lattice2022),\\
  8-13 August, 2022 \\
  Bonn, Germany 
}

\begin{document}
\maketitle

\section{Introduction}

We consider a $\vecp=0$ (spatial momentum zero) correlator 
\begin{equation}
\label{e:Gx0}
	G(t,M,a)= a^3\sum_{\vecx} \langle P^\mathrm{RGI}(x) \overline{P}^\mathrm{RGI}(0) \rangle 
	= G(t,M,0) +\Delta G(t,M,a)\,,
\end{equation}
of a renormalization group invariant (RGI) local field $P^\mathrm{RGI}$ of dimension three
with non-trivial quantum numbers such that the vacuum does
 not contribute as intermediate state. 
 The RGI mass of the theory (or the set of masses) is denoted
 by $M$ and $\Delta O$ denotes the lattice artefact of an observable $O$.
 Weighted integrals $\int G(t,M,0)\, K(t)\, \rmd t$ , such as moments, need a weight 
 $K(t)\simas{t\to0} t^n\,, n>2$ to ensure convergence at small $t$.\footnote{Fields of other dimensions or
 integrals of the type $\int \langle P^\mathrm{RGI}(x) \overline{P}^\mathrm{RGI}(0)\rangle\,\tilde K(x)\, \rmd^4 x$
 lead to trivial changes of our discussion. }
 Specializing to moments with $n>2$, one can then also consider
\begin{equation}
\label{e:Mn}
	\mom_n(M,a)= a\sum_{t} t^n\, G(t,M,a) = \mom_n(M,0) + \Delta \mom_n(M,a)\,,
\end{equation}
with a finite continuum limit $\mom_n(M,0)$. 
The case $n=4$ will be discussed in detail since it is of 
particular interest for computing $\alpha_s$, when $P$ is a heavy-quark bilinear \cite{HPQCD:2008kxl} and furthermore
the hadronic vacuum polarization contribution to $g-2$ of the muon has the form above in the time-momentum representation 
\cite{Bernecker:2011gh} with a $K(t)\simas{t\to0}t^4$. 
We will comment on other moments as we go along. In the following we assume mass-degenerate quarks to simplify the notation.

Note that in the heavy quarks moments method for determining $\alpha_s$ one typically considers the dimensionless 
\begin{equation}
\label{e:Mn}
	\overline{ \mom}_4(M,a)= M^2 \mom_4(M,a) \,,
\end{equation}
with $M$ the RGI-mass, such that also $\overline{ \mom}_4$ is
scale invariant. Specifically one chooses $P^\mathrm{RGI}=Z^\mathrm{RGI} P^\mathrm{bare},\;  P^\mathrm{bare}=\bar c \gamma_5 c'$ and
for discretizations with enough chiral symmetry 
the renormalization factor $Z^\mathrm{RGI}$ is not
 needed due to
$MP^\mathrm{RGI} = m_\mathrm{bare} P^\mathrm{bare}$.
The correlator $G$, \cref{e:Gx0}, is even under time-refections, $G(t,M,0)=G(-t,M,0)$. Thus moments for odd $n$ 
vanish and only moments with $n\geq4$ are finite. 

In an $\rmO(a)$-improved theory, the Symanzik effective theory prediction (SymEFT) \cite{symanzik1983continuum,Husung:2019ytz,Husung:2022kvi} is
\begin{equation}
\label{e:DG}
	\Delta G \simas{a\to 0} a^2 [\alpha_s(1/a)]^{\gammahat_\mathrm{lead}} \,.
\end{equation} 
{\bf Naively} one may expect that this also leads to 
$
	\Delta \mom_n \simas{a\to 0} a^2 [\alpha_s(1/a)]^{\gammahat_\mathrm{lead}}\,.
$
Here we  discuss that this is not the case and show that 
a safe continuum limit cannot even be taken with lattice spacings
down to $a=10^{-2}$fm (\cref{s:demo}). 
We derive that already in the free theory an $a^2 \log(aM)$ term is present (\cref{s:free}) and sketch what changes in the 
 SymEFT prediction in the interacting theory (\cref{s:full}).
 Since the general conclusion is that integrals such as the one
 defining $\mom_4$ cannot be computed reliably on the lattice, we then propose a modification for $\mom_4$ (\cref{s:rho})
 and demonstrate that it works very well. Finally we also make
 a simple and practical proposal which solves the issue for 
 the HVP contribution to the muon $g-2$ (\cref{s:HVP}).

\vspace{3cm}

\section{Demonstration of the deviations from simple $a^2$ scaling}
\label{s:demo}

We computed $\overline{\mathcal{M}}_4$ (and other moments \cite{Chimirri:2022bsu})
in the quenched approximation on ensembles {\tt sft7 - sft4} \cite{Husung:2017qjz} and {\tt q\_b649 - q\_b616} \cite{Chimirri:2022bsu} with lattice spacings $ a = {  0.01\,\mathrm{fm}} \times 2^{n/2}, n=0\ldots 6$, i.e. $   0.01 \,\mathrm{fm} \leq a \leq 0.08\,\mathrm{fm}$. 
The property $MP^\mathrm{RGI} = m_\mathrm{bare} P^\mathrm{bare}$
is guaranteed by using the twisted mass formulation at maximal twist and double insertions of the Pauli term in SymEFT are
avoided by including the Sheikholeslami-Wohlert term \cite{Sheikholeslami:1985ij}
with non-perturbative improvement coefficient \cite{Luscher:1996ug}. 
Further details are given in \cite{Chimirri:latt2022}.

\begin{figure} 
\centering
	\includegraphics[width=11.5cm]{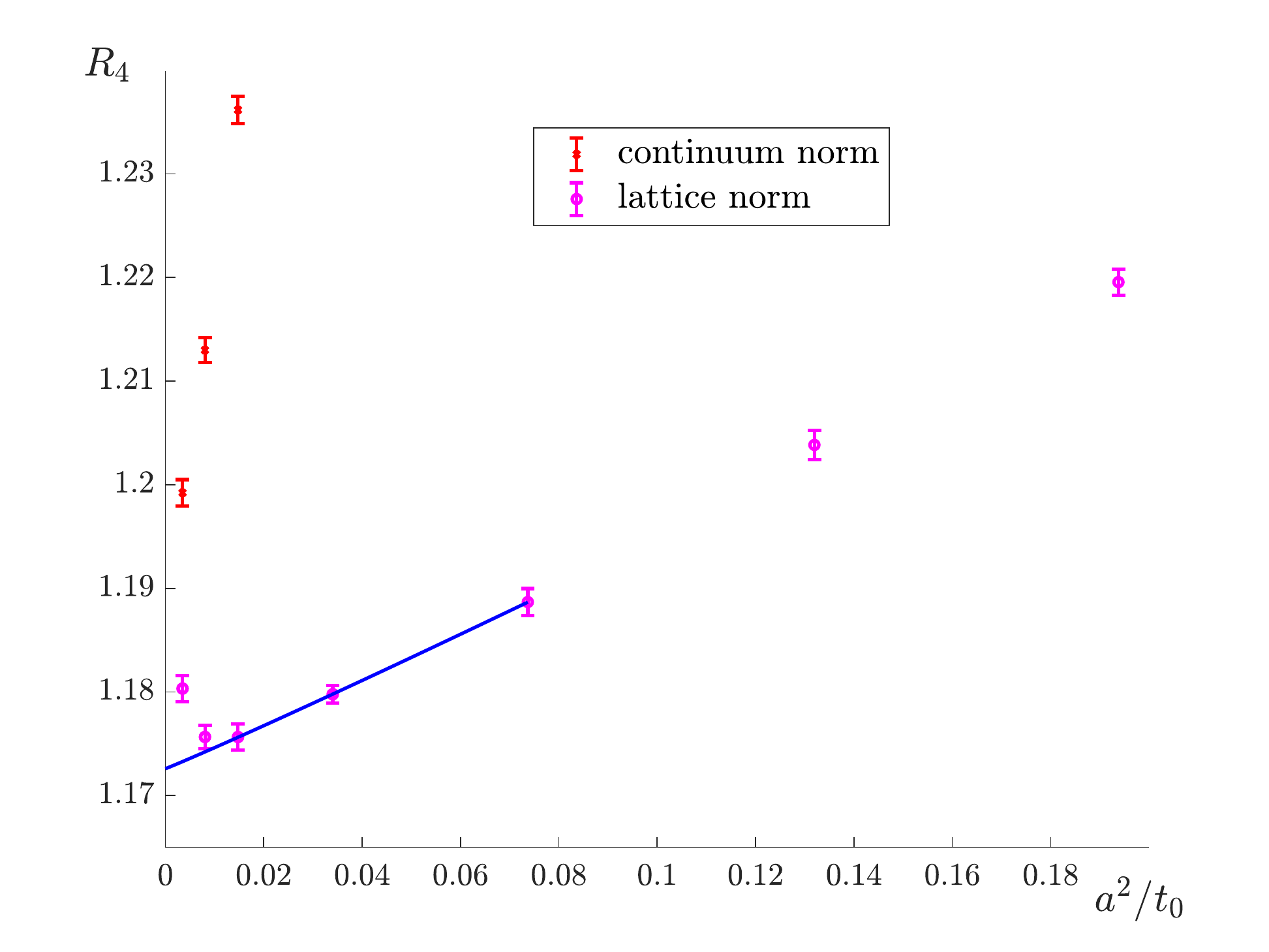} 
	\caption{Lattice dependence of $R_4=\overline{\mathcal{M}}_4 / [\overline{\mathcal{M}}_4]_{g=0}$, where the normalization is
	performed with  $[\overline{\mathcal{M}}_4]_{g=0}$ at finite $a$ (``lattice norm'') or at $a=0$ (``continuum norm''). The quark mass is around the charm quark mass.\label{f:R4}
	}
\end{figure}

In \cref{f:R4} we show the lattice spacing dependence 
of 
\begin{equation}
	R_4=\frac{\overline{\mathcal{M}}_4 }{ [\overline{\mathcal{M}}_4]_{g=0, a>0}}\,.
\end{equation}
The normalization by the lattice leading perturbative order (finite $a>0$) is crucial as seen by the points with continuum norm, $\overline{\mathcal{M}}_4 / [\overline{\mathcal{M}}_4]_{g=0, a=0}$. Again we refer to \cite{Chimirri:2022bsu,Chimirri:latt2022} for more details. However, despite the strong reduction of discretisation errors by the lattice norm, 
a continuum extrapolation with data in the range $a\in [0.02,0.04]$~fm (linear fit in \cref{f:R4}) where $R_4$ seemingly scales with $a^2$ corrections, clearly leads to a wrong result.  
This is seen  by the $a=0.01$~fm data point and corroborated by our method sketched in \cref{s:rho}.
Such a behavior is the nightmare of numerical analysis. Note that the mass $M\approx M_\mathrm{charm}$ is not that high.

\section{Derivation of the $a^2 \log(aM)$ term in the free theory}
\label{s:free}

In this and the following section we study the small $t$ behavior where mass-effects can be neglected and we first consider the contribution to $\mom_4$ from a range 
$t_1\leq t \leq t_2 \ll 1/M$,
\begin{eqnarray}
	\Delta I(t_1,t_2) &=& 2 a\sum_{t=t_1}^{t_2}w_\mathrm{T}(t) \,t^4\, G(t,M,a) - I_\mathrm{cont}(t_1,t_2) \,, \quad t_1M\ll 1, \; t_2M\ll1\,,
	\\
	&& I_\mathrm{cont}(t_1,t_2)=2\int_{t_1}^{t_2} \rmd t \; t^4\, G(t,M,0) \,, \quad t_1M\ll 1, \; t_2M\ll1\,. 	
\end{eqnarray}
The weight $w_\mathrm{T}(t)$ implements the trapezoidal rule: it is $1/2$ 
at the boundaries and $1$ otherwise.

In order to gain understanding, we start with the free theory, $g=0$.
This case is illuminating and at the same time we can get the 
relevant result by dimensional reasoning alone.

We split 
\begin{eqnarray}
\label{e:M2}
 \Delta I(0,t) &=&  \Delta I(0,t_1) +  \Delta I(t_1,t) \,,
\end{eqnarray}
discuss the second term and then add the first one. 
The SymEFT prediction for the cutoff effects of $G$ are 
\begin{eqnarray}
 \Delta G &=& \kl \,\frac{a^2}{t^5}  +\rmO(a^4) +\rmO(M^2t^2) \,,
\end{eqnarray}
with a constant $\kl$ which depends on the fermion discretization.\footnote{This form is simply due to dimensional 
counting. $G(t)$ has mass dimension $-3$ and therefore behaves like $\sim t^{-3}$ for small $t$ in the free theory. In the interacting theory there are log-corrections to that functional form due to anomalous dimensions of $P$ and the SymEFT operators.
Relative cutoff effects are $\sim a^2/t^2$, again because for $tM\ll1$
the only dimensionful parameter apart from $a$ is $t$.}
Performing an explicit leading order computation, expanded
in $a/t$ in the Wilson regularization we find $\kl=1$. Since mass-effects are irrelevant, $\kl=1$ holds irrespective of whether we choose a twisted mass term or a standard one.
Not indicating the higher order corrections in $a$ and $M$ any further we  get
\begin{eqnarray}
 \Delta I(t_1,t)
	 &\overset{a \ll t_1}{\sim}&\; 
	 \kl \,a^2\ \, \int_{t_1}^t \,\rmd s \,s^{-1}  + \Delta I_\mathrm{T}(t_1,t)\,  
	\\ &=& \kl a^2\, \, \log(t/t_1) + \Delta I_\mathrm{T}(t_1,t)  = \kl a^2\, [\log(t/a)-\log(t_1/a)] + \Delta I_\mathrm{T}(t_1,t)\,.
\end{eqnarray}
Here, $\Delta I_\mathrm{T} \sim a^2$ is the error in using the trapezoidal rule for the integral. We drop it because it does not play a role in the following; it is regular as $t\to0$ and does not introduce a log. 
We then obtain
\begin{eqnarray}
\label{eq:DI0t}
  	\Delta I(0,t) &=& \underbrace{\Delta I(0,t_1) -\kl a^2\log(t_1/a)}_{= k\, a^2 } + \kl\,a^2 \,\log(t/a)
  	= a^2\,[k +\kl \log(t/a)] \,,
\end{eqnarray}
with another dimensionless constant $k$ depending on the regularization. The first term, $ka^2$, has this form 
because it neither depends on $t_1$ nor on $t$ and $a$ is the only dimensionful parameter. 

For the full moment, $t$ gets replaced by the only physics scale of the integral, namely $1/M$. 
We thus arrive at
\begin{eqnarray}
  	\frac{\Delta \mom_4(M,a)}{\mom_4(M,0)}  &=& a^2 M^2\,[k' -\kl \log(M\,a)] +\rmO(M^4a^4) \,\,.
\end{eqnarray}

We note that \cite{Harris:2021azd} have argued for the presence of a 
$\log(t/a)$ term in the same discretised integral (in the context of HVP). In contrast to their argumentation, we never work with divergent integrals or with the Symanzik expansion for $a/t =\rmO(1)$. 

It is instructive to add higher order terms, $k_d\,(a^2/t^5)\,(a/t)^{d-2}$ with $ d>2$ terms in the SymEFT for
$\Delta G$.
They yield 
\begin{eqnarray}
\label{e:DItree}
 \Delta I(t_1,t)
	 &\overset{a \ll t_1}{\sim}&\; 
	  \kl a^2\, [\log(t/a)-\log(t_1/a)] +a^2\sum_{d>2} \frac{k_d}{d-2}\,[\,(a/t_1)^{d-2} - (a/t)^{d-2}\,]\,.
\end{eqnarray}
 and
 \begin{eqnarray}
 \Delta I(0,t)
	 &\sim&\; a^2 k''\,+\,
	  \kl a^2\,\log(t/a) - a^2 \sum_{d>2} \frac{k_d}{d-2} (a/t)^{d-2}\,,
\end{eqnarray}
where $k''$ now receives contributions also from the $d>2$ terms in $\Delta G$. Note that the reasoning for the term $a^2k'$ is unchanged. It is simply the dimension of $\Delta I$ inherited from the one of $I$ and the independence on $t_1$. 
This means 
that Symanzik improvement does not hold for the integral: we could improve $\Delta G$ such that all $\sim a^2$ terms are removed, but $\Delta \mathcal{M}_4$ would remain of order $a^2$ due to the $d>2$  terms in \eqref{e:DItree}. "Only" the log-term at order $a^2$ disappears by improvement of the integrand.

Consider for a moment the moment
\begin{equation}
	N_3= a\sum_{t\geq0} t^3\, G(t,M,a) \,.
\end{equation}
In this case, we obtain $\rmO(a)$ effects, irrespective of
how the theory was improved. It is relevant to investigate 
whether such terms appear in some (sub-)integrals in 
representations of light-by-light scattering evaluated on the lattice \cite{Chao:2021tvp}.

\section{SymEFT analysis beyond the free theory}
\label{s:full}
It is not  difficult to follow the above steps for 
the interacting theory. 
One has to write $\Delta G$ as in \cref{e:DG} 
 and also the short distance behavior changes 
due to anomalous dimension effects. 
These modifications introduce powers of $\alpha_s(1/a)$ and  $\alpha_s(1/t)$, respectively, 
but are not
of prime relevance. More important is that
the step analogous to \cref{eq:DI0t} is modified to
\begin{equation}
	k a^2 \to K(a\Lambda)\, a^2\,,
\end{equation}
with a  function $K(a\Lambda)$ which is not restricted by
simple arguments. Without knowing
the behavior of $K$ at the origin, nothing can be concluded 
about $M$-independent $a$-effects of the integral. 
The structure 
of external scale dependent cutoff effects will be discussed 
in a publication \cite{Chimirri:inprep}. The basic 
reason for the difficulty is of course that the interacting theory has 
a dynamical scale, $\Lambda$, which makes the dimensional analysis much less restrictive.

\section{Higher moments $\mom_n,\; n>4$}
\label{s:generaln}
With $n>4$, the $a^2 \log(aM)$ term is absent in the free theory. Still, $\log(aM)$ dependences are present, but they 
are pushed to a higher order in $a$, 
\begin{equation}
  \mom_n = \ldots  + \mathrm{const.}\,\times a^{n-2}\log(aM) + \ldots\,.\qquad 
\end{equation}
%
\section{Solutions}
\label{s:sol}
Our discussion shows that integrals of the considered type cannot be 
computed on the lattice in the straight-forward way. 
The best solution to this problem is to avoid 
integrands which have a behavior $\sim t^k\,, k<2$. First we
describe a specific solution for $\overline \mom_4$ for which we have 
a complete numerical demonstration. Then we propose a
general solution, which in particular will be useful for HVP.

\subsection{A practical solution for $\overline \mom_4$}
\label{s:rho}

Our simple solution for the moment $\overline \mom_4$ uses two different masses in the form
 (dropping the $a$-dependence)
\begin{align}
\label{e:rhodef}
  \rho(M_1,M_2) & =  \frac{2\pi^2}{3}\,(1-r^2)^{-1} \,
  [\overline{\mathcal{M}}_4(M_1) -r^2   \overline{\mathcal{M}}_4(M_2)]
  \\ & =  \frac{2\pi^2}{3}\, (1-r^2)^{-1}  M_1^2 [\mom_4(M_1) -\mom_4(M_2)]\,, \quad r=M_1/M_2 >1 .
\end{align}
The second line shows that the small $t$ asymptotics of the integrand is improved via,
\begin{eqnarray}
	 t^4\,[G(t,M_1)-G(t,M_2)] \sim t^4 \,(t^2 M_2^2- t^2 M_1^2) +\rmO(t^8)\,.
\end{eqnarray}
There are log-corrections to this equation in the interacting theory, which are not relevant here. 
Due to the extra two powers of $t$, which come with the mass-effects, the quantity $\rho(M_1,M_2)$  has no log-enhanced $a^2$ effects (they will appear only at the level
$a^4$). 

For the purpose of extracting $\alpha_s$ it is now 
relevant to choose $M_1$ and $M_2$ not too different. Then
the perturbative expansion, which is given in terms of the one of 
$\overline{\mathcal{M}}_4$, does not contain large logs of
$M_2/M_1$. We write the perturbative expansion in terms 
\footnote{We implicitly define $m_{\star}=\overline{m}_{\msbar}(m_{\star})$. In practice, to evaluate $\overline{\mathcal{M}}_4(M_2)$ we choose $\alpha_{\msbar}( m_{2\star} )$ as expansion variable and use the 5-loop running of the coupling and quark mass to relate $m_{2\star}$ to $M_2$. One could also obtain 
	expansion coefficients which depend on $r$. }
of $\alpha_s(m_{2\star})$ with $M_1>M_2$. The $\frac{2\pi^2}{3}\,(1-r^2)^{-1}$ 
normalization in \cref{e:rhodef} ensures
\begin{equation}
	\rho(M_1,M_2) = 1 + c_1 \alpha_{\msbar}( m_{2\star} )+ \ldots \,,
\end{equation} 
where $c_1=0.74272...$ is the same expansion coefficient as the one of 
$R_4=\frac{2\pi^2}{3} \overline{\mathcal{M}}_4$ and higher order ones are easily obtained. 
We expand in $\alpha_{\msbar}( m_{2\star}) $ because
the difference is dominated somewhat more by long distances
and $M_2$ is the smaller of the masses.

\begin{figure} 
\centering
\includegraphics[width=7.5cm]{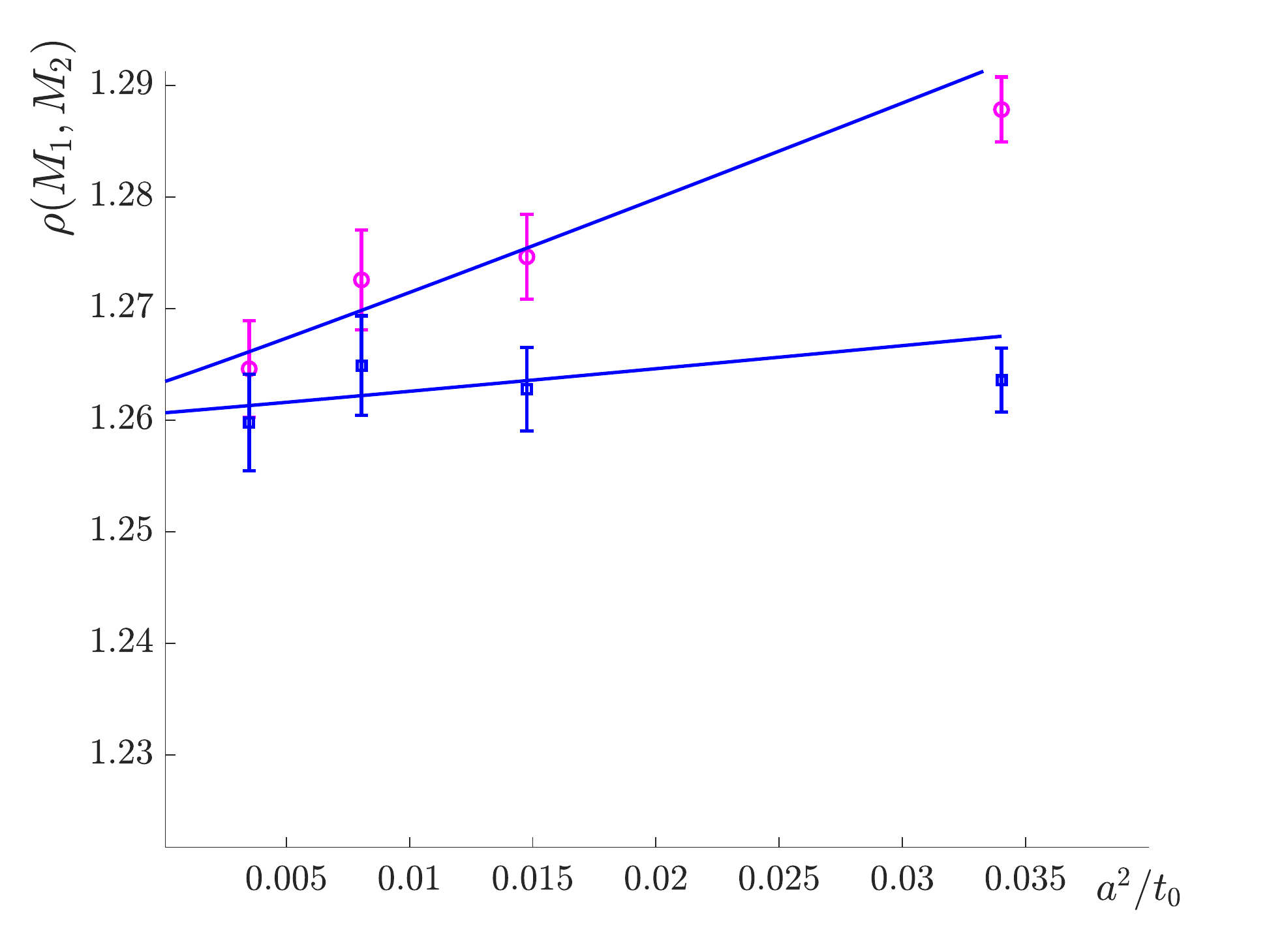} 
\includegraphics[width=7.5cm]{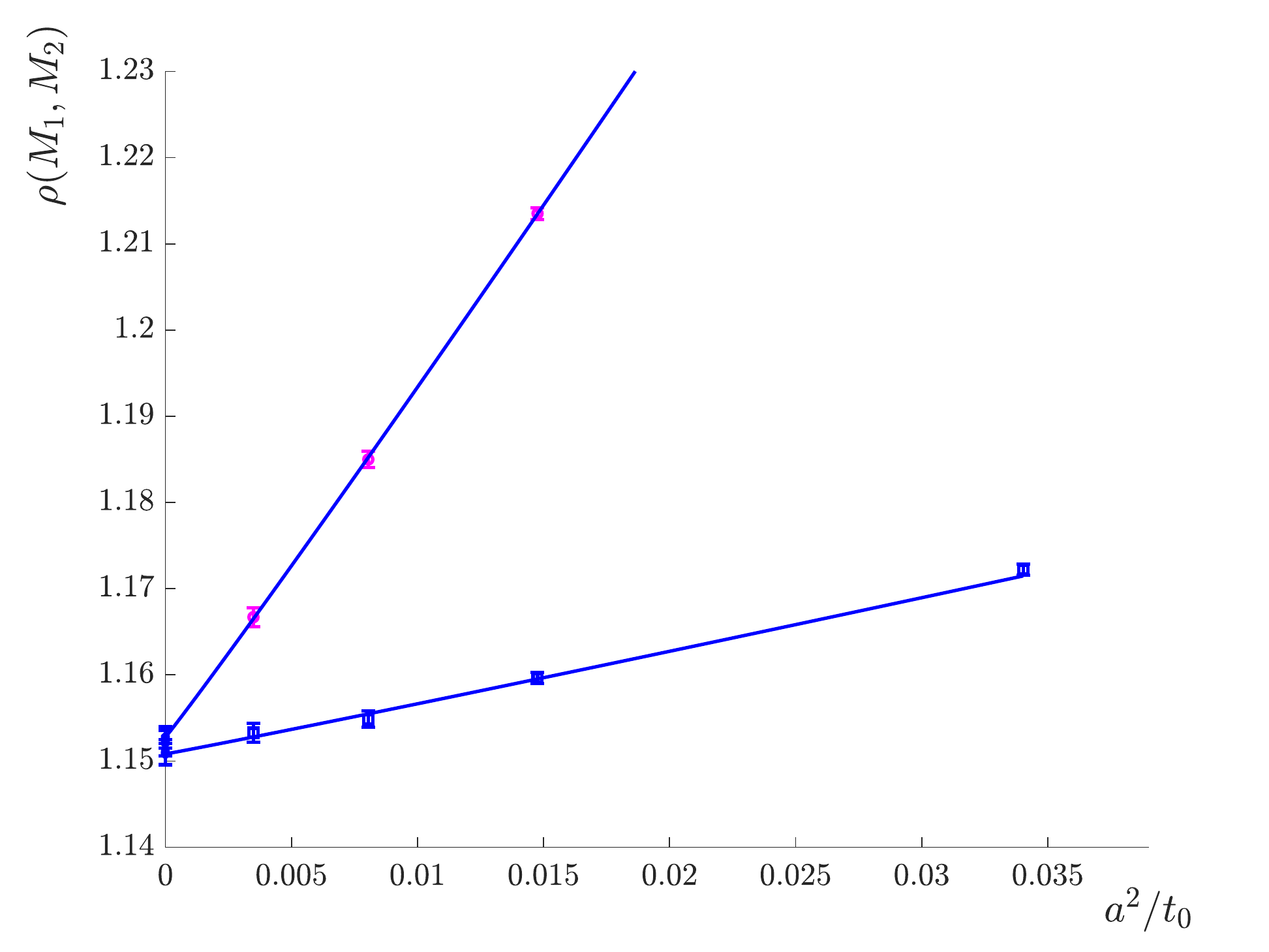} 
	\caption{Continuum limit extrapolations of $\rho$  and its  TL improved version, $\rho(M_1,M_2)^\mathrm{Latnorm}$.  Masses, specified in units
	$z_i=M_i\sqrt{8t_0}$, are $z_1=4.5,\;z_2=3$ (left) and  $z_1=13.5,\;z_2=9$	(right).}
	\label{fig:rho_tl_impr}
	\label{fig:rho}
\end{figure}

We show  continuum limit extrapolations in \cref{fig:rho}. They are almost straight in $a^2$ at small $a$ which makes them quite easy to do. They can be further improved by dividing $\rho$ by the same function 
evaluated at leading order, i.e. $g=0$. There is a choice which masses to insert into the leading order formula. 
A good choice is again $m_\star$. Precisely we define 
\begin{equation}
	R_4^\mathrm{TL}(a\mu) = R_4|_{g=0}
\end{equation}
with $\mu$ the twisted mass and then 
\begin{eqnarray}
  \rho^\mathrm{Latnorm}(M_1,M_2) & = &  \frac{3}{2\pi^2}\,(1-r_\star^2) \,
  \frac{\rho(M_1,M_2)}{R_4^\mathrm{TL}(am_{\star1})-r_\star^2 R_4^\mathrm{TL}(am_{\star2})
  } 
\end{eqnarray}
with
\begin{equation}
	r_\star=\frac{m_{\star1}}{m_{\star2}}\,.
\end{equation}
In principle it is important that $r_\star$ is given by
the ratio of the masses that appear in $R_4^\mathrm{TL}$
for the log-term to cancel.
But numerically, replacing  $r_\star\to r$ makes only a small difference. Examples for how the discretization errors are reduced can be seen in \cref{fig:rho_tl_impr}. For all our
values of $M_1,M_2$, the leading order improved $\rho(M_1,M_2)^\mathrm{Latnorm}$ has a rather convincing continuum extrapolation.

\begin{figure} 
\centering
	\includegraphics[width=10cm]{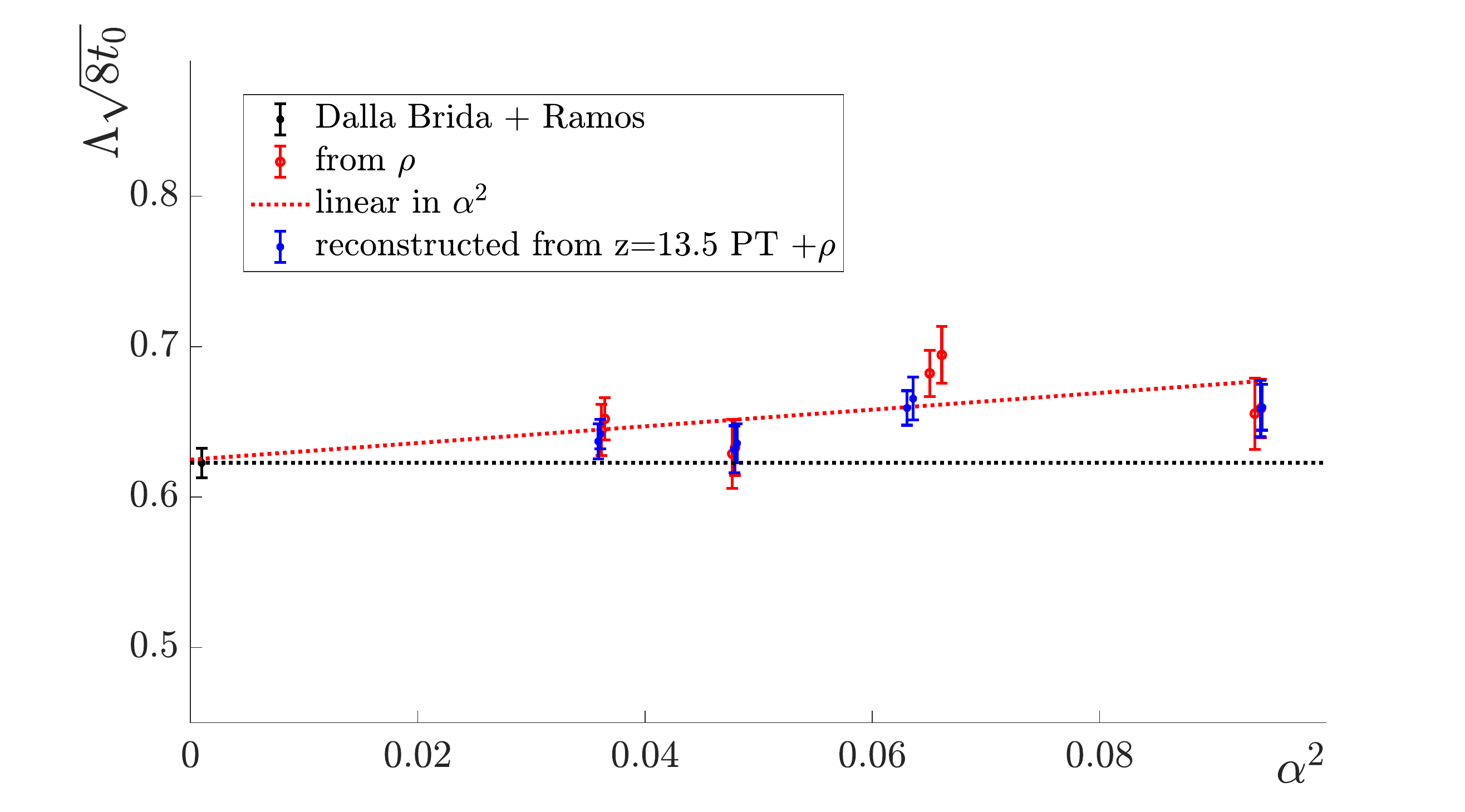}
	\caption{$\Lambda_{\msbar}$ computed from $\alpha_{\msbar}( m_{2\star})$, where the latter is obtained from the non-perturbative $\rho$. The dotted line is a fit to all points including the Dalla Brida / Ramos one \cite{DallaBrida:2019wur}. The reconstructed data points are described in the text.
	}
	\label{fig:lambda}
\end{figure}

After the continuum extrapolation, one  straight-forwardly extracts the effective $\Lambda$-parameter
and arrives at the red circles  in \cref{fig:lambda}. These values  are computed
from three-loop perturbation theory (i.e. including $\alpha^3$ in $R_4$) at finite $\alpha(m_\star)$. They then have a residual
dependence 
\begin{equation}
	\Lambda_\mathrm{eff} = \Lambda +\rmO(\alpha^2(m_\star))\,,
\end{equation}
on $m_\star$ and we call them ``effective''. 
The comparison to the  Dalla Brida and Ramos 
value \cite{DallaBrida:2019wur},
extracted at $\alpha^2 < 0.01$ with the help of a finite size step scaling method, shows that $\Lambda$ computed from 
$\rho$ has at most small (on the scale of our uncertainties) corrections at the largest mass. That mass is given by $z=9$ or $ m_\star\approx 2.7\,\mathrm{GeV}$.

\subsection{Reconstruction of $R_4=\frac{2\pi^2}{3}\overline{\mathcal{M}}_4(M)$ from $\rho$.}

From the definition \cref{e:rhodef}
of $\rho$ it is clear that 
given $\rho(M_1,M_2)$ and $R_4(M_2)$ one can determine $R_4(M_1)$. This can be exploited by using $\rho$ to go from $R_4(M_\mathrm{ref} \gg \Lambda)$, where perturbative uncertainties are suppressed the most, to smaller
masses.\footnote{In the opposite direction {\em all} uncertainties in $\rho$ get enhanced, quickly leading to uncontrolled results.} We insert the known \cite{DallaBrida:2019wur} $\Lambda$-parameter into the three-loop (i.e. including $\alpha^3$) perturbative expression for $R_4$
at our highest mass,
$z_\mathrm{ref}=13.5$ and obtain
\begin{eqnarray}
	R_4^\mathrm{reconstructed}(M)  &=& (1-r^{-2}) \, \rho(M,M_\mathrm{ref}) + r^{-2} \,R_4^\mathrm{3-loop}(M_\mathrm{ref}) \,,\; 
	\\ && r=M/M_\mathrm{ref}, \; z_\mathrm{ref}=\sqrt{8t_0}M_\mathrm{ref}=13.5	\,.
\end{eqnarray}
Note that perturbative errors are small in
$R_4(M_\mathrm{ref})$ as seen in the analysis of $\rho$. They get further suppressed by a factor $r^{-2}\approx 1/20$ when we go to
$z=3$. This means that we  obtain the non-perturbative dependence of  $\Lambda_\mathrm{eff}$ (as of now computed from $R_4$ and therefore with somewhat different 
$\rmO(\alpha^2)$ terms) on 
$\alpha$. We remind the reader that a direct computation of $R_4$ was impossible due to the $a^2 \log(aM)$ effects.

\subsection{Proposal for the HVP contribution to the muon $g-2$}
\label{s:HVP}
The discussion in the previous section is easily
transferred to the case of the muon $g-2$, working with differences of the HVP integral
for different (artificial) muon masses. Additionally,
we would like to advocate a very simple solution 
for this and similar cases, where 
the short distance contribution to the integral is subdominant. In contrast to the $\mom_4$-case
the goal is {\em not} to determine $\alpha_s$ or other short-distance parameters. 

It is then advisable to split the integral into a
short-distance part evaluated by continuum perturbation theory and a long-distance one to be computed on the lattice:
\begin{equation}
	 \int_0^\infty \mathrm{d} t F(t)=  \underbrace{\int_0^\infty \mathrm{d} t\,[1-\chi(t)]\,F(t)}_{\text{continuum PT} }\,+\,  
	 \underbrace{a\sum_{t=0}^\infty\,\chi(t)\,\,F(t)}_{\text{continuum limit of lattice results}}  \,,\quad   \chi(t) \,\sim\,\begin{cases}\mathrm{O}(t^2) & t\Lambda_\mathrm{\overline{MS}}\ll1 \\ 1 &t\Lambda_\mathrm{\overline{MS}} \gg 1\end{cases}\,.
\end{equation} 
For example the function $\chi$
can be taken as 
\begin{equation}
\label{e:chi}
\chi(t) = \frac{(M_\mathrm{cut} t)^k}{(M_\mathrm{cut} t)^k  + 1}\,, \; M_\mathrm{cut}\gg \Lambda_\mathrm{\overline{MS}}
\end{equation} 
or also as a step-function, $\chi(t)=\theta(t M_\mathrm{cut} -1)$. 
The smooth version seems advantageous for perturbation theory 
as well as for the lattice discretization of the integral. 
The use of perturbation theory 
for the small $t$-part of the integral has already been 
anticipated in \cite{Bernecker:2011gh}. Our discussion adds
further motivation and understanding. It suggests a smooth function 
$\chi$ such as \cref{e:chi}.
\\[1ex]
{\bf Acknowledgements.} We thank the organizers for a very pleasant and successful conference.  Discussions  with H. Meyer, S. Kuberski and A. Risch on HVP are gratefully acknowledged. We also thank
C. Lehner for an email exchange on that subject. LC and RS acknowledge funding from the European Union’s Horizon 2020 research and
innovation programme under the Marie Skłodowska-Curie grant agreement No. 813942.

\bibliographystyle{./JHEP}
\bibliography{./refs}

\end{document}